\documentclass[epj]{svjour}
\usepackage{graphics}
\usepackage{graphicx}
\usepackage{color}

\newcommand{\vecr}{\mathbf{r}}

\newcommand{\veca}{\mathbf{a}}

\newcommand{\vecF}{\mathbf{F}}

\newcommand{\vecv}{\mathbf{v}}

\newcommand{\Wcm}{\;\mathrm{W/cm}^2}

\newcommand{\beq}{\begin{equation}}
\newcommand{\eeq}{\end{equation}}

\begin{document}
\title{Quantum effects on radiation friction driven magnetic field generation}
\author{Tatyana V. Liseykina\inst{1,2,3} \and Andrea Macchi\inst{4,5} \and Sergey V. Popruzhenko\inst{6,7}\\
\email{sergey.popruzhenko@gmail.com}
}                     
%
%
\institute{Institute of Physics, University of Rostock, 18051 Rostock, Germany \and Institute of Computational Mathematics and Mathematical Geophysics SD RAS, 630090 Novosibirsk, Russia \and Nikolski Institute of Mathematics (RUDN), 117198 Moscow, Russia \and CNR, National Institute of Optics (INO), Adriano Gozzini research unit, Pisa, Italy \and Enrico Fermi Department of Physics, University of Pisa, 56127 Pisa, Italy \and Prokhorov  General Physics Institute RAS, 119991 Moscow, Russia \and Institute of Applied Physics RAS, 603950 Nizhny Novgorod, Russia}

\date{Received: date / Revised version: date}
%
\abstract{
Radiation losses in the interaction of superintense circularly polarized laser pulses with high-density plasmas can lead to the generation of strong quasistatic magnetic fields via absorption of the photon angular momentum (so called \emph{inverse Faraday effect}). 
To achieve the magnetic field strength of several Giga Gauss laser intensities $\simeq 10^{24}\Wcm$ are required which brings the interaction to the border between the classical and the quantum regimes.
We improve the classical modeling of the laser interaction with overcritical plasma in the ``hole boring'' regime by using a modified radiation friction force accounting for quantum recoil and spectral cut-off at high energies. 
The results of analytical calculations and three-dimensional particle-in-cell simulations show that, in foreseeable scenarios, the quantum effects may lead to a decrease of the conversion rate of laser radiation into high-energy photons by a factor 2--3.
The magnetic field amplitude is suppressed accordingly, and the magnetic field energy -- by more than one order in magnitude.
This quantum suppression is shown to reach a maximum at a certain value of intensity, and does not grow with the further increase of intensities.
The non monotonic behavior of the quantum suppression factor results from the joint effect of the longitudinal plasma acceleration and the radiation reaction force.
The predicted features could serve as a suitable diagnostic for radiation friction theories. 
} 
\maketitle
\section{Introduction}
\label{sec:intro}
The next generation of high power laser systems \cite{danson-hpl15,light-f,ELI,P3-ELI,apollon,J-Karen,SULF,SULF-1,SULF-2,APRI,CAEP,XCELS} will lead to a new regime of laser-plasma interactions where the emission of incoherent radiation in the X- and $\gamma$-ray range plays a dominant role in the plasma energetics (see e.g. \cite{bulanov-rmp06,dipiazza-rmp12,fedotov-cp15} for review). 
The modeling of these conditions requires the essential inclusion of radiation friction (RF) effects, i.e. to account self-consistently for the effect of radiation emission on the electron dynamics.  
In the classical context, RF is accounted for by introducing an additional term to the Lorentz force, acting effectively as friction whose work equals the energy lost into radiation. 
The expression for the classical RF force has been the subject of a longstanding controversy and several different forms have been proposed. 
However, it now appears that the expression proposed by Landau and Lifshitz \cite{landau-lifshitz-RR} is accurate enough as far as one can neglect quantum effects. 
The latter become important when the individual energy and momentum of the emitted photons are not small with respect to the energy and momentum of the emitting electron. 
Although in principle an exact description of the self-consistent interaction of an electron with a strong electromagnetic field may be provided by quantum electrodynamics (QED), in practice such description is technically unfeasible and approximate models are needed. 

Recent experiments based on Thomson scattering of a superintense laser pulse by a high-energy electron bunch \cite{cole-prx18,poder-prx18,macchi-p18} have provided a first (albeit weak) evidence for deviations from classical predictions in the radiation spectrum. 
In the modeling of the experiments, a so-called ``semiclassical'' approach based on a modification of the Landau-Lifshitz RF force \cite{kirk-ppcf09} appeared to provide a better agreement than a ``quantum'' approach
where the radiation emission is described stochastically, i.e. considering the emission of individual photons with a given probability distribution. 
This finding has been explained by the possible breakdown of approximations underlying the quantum model, as also suggested by a different experiment \cite{wistisen-natcomm18}, and attempts are being pursued to overcome the approximations in the quantum model \cite{blackburn-pop18,dipiazza-pra18,ildertonPRA19,aleksandrovPRD19,dipiazza-pra19,seiptPPCF19,wistisenPRR19,dinuPRD20}. 
This scenario suggests that identifying different RF signatures and test their sensitivity to the onset of quantum effects is important in order to improve the theoretical and numerical modeling.

In a recent paper \cite{liseykina-njp16}, we showed that RF induces a specific form of the \emph{inverse Faraday effect} (IFE), i.e. the generation of magnetic fields due to absorption of angular momentum into a plasma. 
From a classical viewpoint, in the presence of dissipative effects (e.g. friction) the electromagnetic (EM) angular momentum carried by a circularly polarized laser pulse is transferred to the plasma electrons which acquire a torque and in turn produce an azimuthal current and an axial magnetic field.
From a quantum viewpoint, the RF-induced absorption of angular momentum is due to the annihilation of $N\gg1$ polarized laser photons needed to generate a single high-energy photon, so that an angular momentum amount equal to $(N-1)\hbar \approx N\hbar$ (being $\hbar$ the photon spin, independent of the photon energy) is transferred to the orbital motion of the electrons.  
In a regime of interaction with high-density plasmas the conversion efficiency $\eta_\mathrm{rad}$, see Eq.~(\ref{eta-1}) for the definition, of laser energy into incoherent radiation  may be a few ten percent, which results in axial quasistatic magnetic fields up to gigagauss values at intensities $\simeq 10^{24}\Wcm,$ as observed in three-dimensional (3D) particle-in-cell (PIC) simulations with classical RF included \cite{liseykina-njp16}.
Such huge magnetic fields, besides affecting the plasma dynamics, provide an unambiguous signature of RF effects and may be measured by polarimetry methods.
  
By further developing the classical model for the calculation of radiation losses \cite{poprz-njp19} we introduced a self-consistent picture of the IFE, accounting for RF effects on electron motion via the Zeldovich model \cite{zeld-spu75}, for the plasma motion driven by radiation pressure (``hole boring'') \cite{macchi-prl05,robinson-ppcf09}, and for the inhomogeneous distribution of the laser intensity. 
This improved model predicts values of $\eta_\mathrm{rad}$ in good agreement with the PIC simulations. 
However, the range of laser intensities considered extends to values high enough for quantum RF effects to become important and possibly dominant.

In this paper, we include quantum effects in our theory and simulations using a semiclassical approach based on the modification of the RF force via the factor introduced by Ritus \cite{ritus-jslr85,ritus}, as was done in the interpretation of experiments \cite{poder-prx18} and in other recent theoretical works \cite{kirk-ppcf09,niel-ppcf18}. 
We find that quantum effects lead to a considerable reduction both in $\eta_\mathrm{rad}$ and in the magnetic field amplitude. 
The suppression factor in the value of the conversion efficiency is found to have a minimum $\mathrm{ SF}\approx 1/2 - 1/3$ at intensity $\approx 10^{24}\Wcm$ indicating the existence of an extended intensity domain where the quantum effects make a quantitatively considerable impact, which however does not grow with intensity because of the interplay of two (detailed below)
opposite tendencies.
This implies that the axial magnetic field remains of the same order of magnitude but its exact value appears quite sensitive to the quantum modification of the RF force.

The paper is organized as follows.
In the section \ref{sec:IFE-classical} we give a brief summary of the IFE theory in ultra relativistic laser plasma as considered in Refs.\cite{liseykina-njp16,poprz-njp19}.
In Section \ref{sec:IFE-semiclassical}, the semiclassical modification of the emission spectrum is explained and the analytic model of \cite{poprz-njp19} is modified accordingly.
Section \ref{sec:results} presents results of PIC simulations where the quantum suppression of the RF force is introduced within the same approach as in the model of Section \ref{sec:IFE-semiclassical}.
The last section contains summary and outlook.

\section{Selfconsistent theory of inverse Faraday effect in the classical regime of interaction}
\label{sec:IFE-classical}
Based on the equations of macroscopic electrodynamics and conservation laws, a description of IFE in the field of an intense laser pulse \cite{liseykina-njp16} predicts the maximal amplitude of the quasistatic longitudinal magnetic field excited on the axis of a laser beam to be proportional 
to the laser magnetic field amplitude $B_L$ and to the fraction of the laser energy $\eta$ associated with the irreversible transfer of angular momentum from the laser field to the plasma:
\beq
B_\mathrm{xm}=C\eta a_0B_0\equiv C\eta B_L~.
\label{Bmax}
\eeq
Here
\beq
a_0=\frac{eE_L}{mc\omega}\equiv\frac{E_L}{B_0}~, ~~~~~B_0=\frac{mc\omega}{e}=1.34\cdot 10^8\mathrm{G}~
\label{a0}
\eeq
are the dimensionless laser field amplitude and the characteristic magnetic filed correspondingly; $E_L=B_L$ and $\omega$ is the laser frequency; the value of $B_0$ in (\ref{a0}) corresponds to the wavelength $\lambda=2\pi/\omega=800$nm of a Ti:Sa laser.
A dimensionless coefficient $C$ is determined by the shape of the laser pulse and has typical values $C\simeq 0.1\div 0.2$.
The structure of Eq.~(\ref{Bmax}) is consistent with the general theory of IFE \cite{pitaevskiiJETP61,haines-prl01}.
The factor $\eta$ is defined as
\beq
\eta=\frac{\omega L_\mathrm{abs}}{U_L}~,
\label{eta}
\eeq 
where $L_\mathrm{abs}$ is the angular momentum absorbed by the plasma and
\beq
U_L={\cal A}\lambda^3a_0^2B_0^2~
\label{UL}
\eeq
is the energy stored in the laser pulse of wavelength $\lambda$.
The dimensionless coefficient $\mathcal{A}$ is determined by the pulse focusing and time envelope.
Equations (\ref{Bmax})--(\ref{eta}) are insensitive to a physical mechanism of the angular momentum transfer.
In particular, Eq.~(\ref{Bmax}) applies independently on the impact of quantum effects on the plasma dynamics.

In the high-field regime, radiation of plasma electrons is the only mechanism for energy dissipation, so that $\omega L_\mathrm{abs}=U_\mathrm{rad}$, where $U_\mathrm{rad}$ is the  energy radiated out by the electrons \cite{liseykina-njp16}, and Eq.~(\ref{eta}) reads
\beq
\eta\equiv\eta_\mathrm{rad}=\frac{U_\mathrm{rad}}{U_L}\equiv\frac{\int d^3r\int dt P_\mathrm{rad}({\bf r},t)n_e({\bf r},t)}{U_L}\le 1~.
\label{eta-1}
\eeq
Here $P_\mathrm{rad}$ is the emission power for a single electron moving under the action of the local electromagnetic field which includes also that created by the plasma and $n_e$ is the electron concentration.
Small values of the characteristic wavelengths of emitted radiation, $\lambda\simeq\lambda_L/a_0^2\simeq 10^{-9}\div 10^{-8}$cm guarantee that the emission process is entirely incoherent, so that possible effects of interference are discarded in Eq.~(\ref{eta-1}).
In order to proceed with the estimation of $\eta_\mathrm{rad}$ 
we adopted the following assumptions \cite{poprz-njp19}:
\begin{enumerate}
\item[(i)] plasma electrons radiate independently in the field of a plane CP electromagnetic wave;
\item[(ii)]the laser field attenuation inside the plasma and the time-space distribution of the laser energy are accounted for while calculating the integral in Eq.~(\ref{eta-1});
\item[(iii)] effect of the RF force on the electron motion is taken into account selfconsistently using the Zeldovich model for a CP electromagnetic wave propagating through a homogeneous plasma \cite{zeld-spu75};
\item[(iv)] the global motion of the plasma slab is accounted for within the ``hole boring'' model \cite{macchi-prl05,robinson-ppcf09};
\item[(v)] effect of quantum recoil on motion and radiation of electrons is discarded.
\end{enumerate}
Despite of simplifications (i)--(iv), the analytic model of \cite{poprz-njp19} reproduces results of 3D 
PIC simulations with a 20\% accuracy\footnote{Such a good quantitative agreement is most probably occasional. A simple analytic model as introduced in \cite{poprz-njp19} is only expected to correctly predict the tendencies and provide order-of-magnitude estimates, see results and discussion in Sections ~\ref{sec:results},\ref{sec:discussion}.} and shows that longitudinal acceleration of the plasma and attenuation of the laser field on a small evanescence length inside the plasma, together with effects of time-space averaging, lead to a considerable suppression in the conversion efficiency $\eta_\mathrm{rad},$ Eq. ~(\ref{eta-1}), so that, ultimately it does not exceed $\eta_\mathrm{rad}\approx 0.2$ for $a_0=600$, leading to the upper limit estimate of the magnetic field, Eq.~(\ref{Bmax}), $B_{xm}\approx 0.04a_0B_0\approx 3.2\cdot 10
^9$G at laser intensities $I_L\approx 1.7\cdot 10^{24}\Wcm.$

As far as restriction (v) is concerned, both the model of \cite{liseykina-njp16,poprz-njp19} and the PIC simulation are based on the classical equations of motion for charged particles and classical expressions of the radiation power discarding the effect of quantum recoil on the spectrum of emitted radiation.
The parameter $\chi$ 
\beq
\chi=\frac{e\hbar}{m^3c^4}\sqrt{-(F_{\mu\nu}p^{\nu})^2}=\frac{E_L^{\prime}}{E_\mathrm{cr}}
\label{chi}
\eeq
which determines the significance of quantum effects and equals to the ratio of the external (laser) electric field in the electron rest frame $E_L^{\prime}$ to the critical field of quantum electrodynamics \cite{sauter,heisenberg,schwinger}, $\displaystyle E_\mathrm{cr}=\frac{m^2c^3}{e\hbar}=1.32\cdot 10^{16}$V/cm, remains smaller than unity up to intensities $I_L\sim 10^{25}\Wcm$ making classical description of dynamics and radiation of electrons applicable at least on the qualitative level.
From the other side, the spectrum of emitted photons appears considerably modified by quantum effects already for $\chi\approx 0.1$ \cite{ritus-jslr85,kirk-ppcf09,niel-ppcf18}.
This is achieved, for the considered parameters, already at $a_0\approx 200$ ($I_L\approx 1.9\cdot 10^{23}\Wcm$), so that quantum corrections to the power of radiation may become numerically significant.
In the following, we qualitatively probe possible manifestations of the quantum effects in the considered problem at intensities which leave the parameter $\chi < 1.$

\section{Quantum effect on the conversion efficiency}
\label{sec:IFE-semiclassical}
First, we introduce the quantum factor $g(\chi)$ which describes the suppression of the radiation power due to the off-set in the emission spectrum \cite{ritus-jslr85,kirk-ppcf09}.
The classical radiation power for a particle moving along a circle with a transverse velocity $v_0$ and drifting in the longitudinal direction with a velocity $v_x$ is given by
\beq
P_\mathrm{rad}=\frac{2e^2\omega^2\gamma^4v_0^2}{3c^3}\bigg(1-\frac{v_x}{c}\bigg)^2\equiv \frac23\alpha\chi^2\frac{m^2c^4}{\hbar}~.
\label{P-1}
\eeq  
To account for the quantum effect of suppression, this classical equation is replaced by 
\beq
\tilde{P}_\mathrm{rad}=g(\chi)P_\mathrm{rad}~.
\label{P-g}
\eeq
General formulas for the quantum factor $g(\chi)$ can be found in \cite{ritus-jslr85,kirk-ppcf09}.
For practical calculations we use a fit suggested in \cite{thomas-prx12}:
\beq
g(\chi)=\bigg(1+12\chi+31\chi^2+3.7\chi^3\bigg)^{-4/9}.
\label{g-chi}
\eeq
In the ultrarelativistic limit 
the RF force is proportional to the radiation power 
\beq
\vecF_\mathrm{rad}=-\frac{P_\mathrm{rad}\vecv}{c^2}~,
\label{RRF-rel}
\eeq
so that the quantum factor enters it in the same way $\vecF_\mathrm{rad}\to\tilde{\vecF}_\mathrm{ rad}=g\vecF_\mathrm{rad}$.
As a result, the equation connecting the gamma-factor $\gamma$ with the dimensionless field amplitude $a_0$ is modified as \cite{zhang-njp15}
\beq
a_0^2=\gamma^2(1+g^2(\chi)\xi^2\gamma^6)~.
\label{a0-gam-g}
\eeq
This equation stems from the equilibrium condition for a stationary circular orbit for an electron moving under the action of the Lorentz and the RF forces, cf. Eq.~(14) in \cite{poprz-njp19}.
The conversion efficiency is then modified accordingly
\beq
\eta_\mathrm{rad}=g(\chi)\xi\frac{\gamma^4}{a_0}~,
\label{eta-g}
\eeq
cf. Eq.~(16) in \cite{poprz-njp19}.
This equation relates to the case of a laser field homogeneous both in the propagation direction and the polarization plane.
Note that in Eqs.~(\ref{a0-gam-g}) and (\ref{eta-g}) the gamma factor $\gamma$ and the parameter 
\beq
\xi=\frac{4\pi r_e}{3\lambda}~, \, \mathrm{with}\,~~r_e=\frac{e^2}{mc^2}
\label{xi}
\eeq
are taken in the reference frame co-moving with the slab of radiating electrons.
In the following, we will denote this reference frame $K$ to distinguish it from the laboratory frame $K_0$ where the corresponding values are denoted as $\gamma_0$ and $\xi_0$.
The same applies to Eq.~(\ref{Bmax}).
Taking into account that a Lorentz transformation does not change the values of electric and magnetic fields along the boost axis, Eq.~(\ref{Bmax}) gives also the magnetic field in the laboratory frame $K_0$, although the conversion efficiency, Eq.~ (\ref{eta-g}), and the values of $B_0,$ Eq.~(\ref{a0}), and $B_L$ are defined exclusively in $K$.
To connect different values in $K_0$ and $K$, we follow the model of \cite{poprz-njp19} and assume below that the average drift velocity of the electrons is close to that of the ions and given by the hole boring velocity \cite{robinson-ppcf09}
\beq
v_x\approx v_\mathrm{HB}=c\frac{\sqrt{\Theta}}{1+\sqrt{\Theta}}~,~~~\Theta=\bigg(\frac{a_0}{a_\mathrm{ HB}}\bigg)^2
\label{vHB}
\eeq
with the parameter
\beq
a_\mathrm{HB}=\sqrt{\frac{An_0m_p}{Zn_cm}}~,
\label{aHB}
\eeq
where $A$ and $Z$ are the atomic and charge numbers of the ions, $n_0$ is the initial electron concentration, $n_c=m\omega^2/4\pi e^2$ is the critical concentration and $m_p$ is the proton mass.
In the present paper the simulation was performed for $Z/A=1/2$ and $n_0=90n_c=1.55\cdot 10^{23}\mathrm{cm}^{-3}$ (same as in \cite{liseykina-njp16,poprz-njp19}), then $a_\mathrm{HB}\approx 6\cdot 10^2$.
The longitudinal motion becomes significantly relativistic at $a_0\simeq a_\mathrm{HB}$.

Using Zeldovich solution \cite{zeld-spu75} the quantum parameter $\chi$ 
can be expressed via the values $\xi$ and $\gamma$ in the frame K as \cite{poprz-njp19}
\beq
\chi=\frac{3\hbar c}{2e^2}\xi\gamma^2\approx 2\cdot 10^2\xi\gamma^2~.
\label{chi-xi}
\eeq
Within this model,  the values of $\gamma$ and, consequently, of $\eta_\mathrm{rad}$ are determined by two parameters $a_0/a_\mathrm{HB}$ and $a_0/a_\mathrm{cr}$ with 
\beq
a_\mathrm{cr}=\xi_0^{-1/3}\approx 400~,
\label{acr}
\eeq
where $\xi_0$ is given by Eq.~(\ref{xi}) calculated in the frame $K_0$ for $\lambda_0=800$nm.
For a flat-top laser pulse with intensity homogeneously distributed in space and time, the conversion efficiency is given by the ratio of the full radiation power of all the electrons to the laser intensity.
A simple calculation (see \cite{poprz-njp19} for details) gives then 
\beq
\eta_\mathrm{rad}=\frac{2\xi g(\chi)a_0^3f(a_0,\xi,\chi)}{1+v_\mathrm{ HB}(a_0)/c}~,~~~f(a_0,\xi,\chi)=\frac{1}{a_0^5}\int\limits_0^{a_0}\gamma^4(a',\xi,\chi)da'~,
\label{eta-2}
\eeq
where the function $f$ accounts for the laser field attenuation in the plasma.
Finally, the fully consistent definition, Eq. ~ (\ref{eta-1}), involves the ratio of emitted to laser energy which requires integration of the emission power over time and radial coordinate.
This definition applies also for laser pulses with an arbitrary space-time envelope and leads (cf. Eqs.~(31), (32) in \cite{poprz-njp19}) to the expression 
\beq
\eta_\mathrm{rad}=\frac{2}{a_0^2}\cdot\frac{\int\limits_0^{\infty}d\rho\int\limits_0^{\infty}d\tau[ \xi(\rho,\tau)g(\rho,\tau)/(1+v_\mathrm{ HB}(\rho,\tau)/c)]\int\limits_0^{a_0F(\rho,\tau)}\gamma^4(a,\xi,\chi)da}{\int\limits_0^{\infty}d\rho\int\limits_0^{\infty} F^2(\rho,\tau)d\tau~}.
\label{eta-3}
\eeq
Here $F(\rho,\tau)$ is the axially symmetric laser pulse space-time envelope defined as
\beq
a(r,t)=a_0F(\rho,\tau)
\label{F}
\eeq
with dimensionless radial and time variables $\rho=(r/r_0)^2$, $\tau=ct/r_L$ and
\beq
\xi(\rho,\tau)=\xi_0\sqrt{\frac{1-v_\mathrm{HB}/c}{1+v_\mathrm{HB}/c}}~
\label{xi-a0}
\eeq
is calculated in the frame $K$ with the hole boring velocity $v_\mathrm{HB}$ dependent on time and radial coordinate through the value of the laser field amplitude, Eq.~(\ref{F}).
The $\gamma$-factor under the integral in the numerator depends on $\xi(a_0),$ 
Eq.~(\ref{a0-gam-g}), and also on the local value of the dimensionless laser amplitude $a_0F(\rho,\tau)$.
Note that the quantum factor $g(\chi)$ enters Eq.~(\ref{eta-3}) not only linearly but also through the factor $\gamma^4$ under the integral.
Finally, the factor $1/(1+v_\mathrm{HB}/c)$ in Eqs.~(\ref{eta-2}) and (\ref{eta-3}) comes from the Lorentz transformation of the emitted energy into the laboratory frame $K_0$.

\section{Numerical results}
\label{sec:results}
\begin{figure*}
\centering
\includegraphics[width=0.9\textwidth]{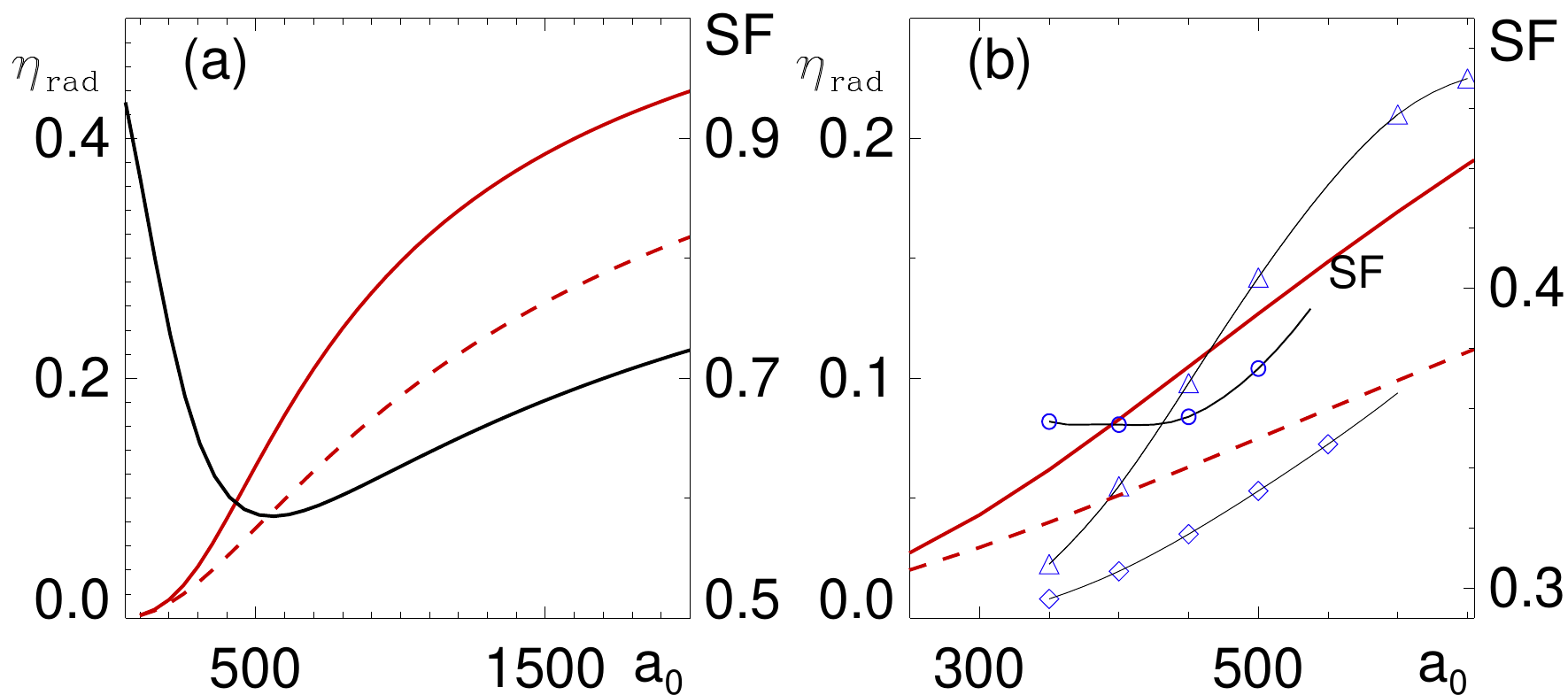}
\caption{Conversion efficiency $\eta_\mathrm{rad}$ calculated with $g=1$ (solid red lines) and $g(\chi)$ from Eq.~(\ref{g-chi}) (dashed red lines).
The panel (a) shows results of the analytic model of Sec.~\ref{sec:IFE-semiclassical} up to $a_0=2\cdot 10^3$, the panel (b) zooms into the interval $a_0\in [250,650].$ The PIC simulation have been performed for $a_0\in [350,650].$
PIC results for $\eta_\mathrm{rad}$ are shown by triangles for $g=1$ and by diamonds for $g=g(\chi)$.
The suppression factor SF, Eq.~(\ref{SF}), calculated from the theory (thick black line) and extracted from the PIC data (circles) reaches the minimal values $\approx 0.58$ and $\approx 0.36$, correspondingly.}  
\label{fig1}
\end{figure*}

By solving numerically Eq.~(\ref{a0-gam-g}) using Eqs.~(\ref{g-chi}) and (\ref{chi-xi}) for the quantum factor $g$  and  the quantum parameter $\chi,$ correspondingly, the conversion efficiency, Eqs.~(\ref{eta-2}), (\ref{eta-3}), 
has been found in the interval of intensities $10^{23} -- 10^{25}\Wcm.$
Fig.~\ref{fig1} shows the conversion efficiency $\eta_\mathrm{rad}$ calculated in the frame $K_0$ with and without the function $g(\chi)$ accounted for.
Note that although for an electron which is in average in rest in the frame $K_0$ the quantum factor $g$ would change from 0.8 at $a_0=100$ to 0.2 at $a_0=1000$, and $\chi\approx 3$ at the upper limit of this integral, the actual effect on $\eta_\mathrm{rad}$ does not grow monotonically but achieves a maximum in the interval $a_0=300 - 600$ where the two curves calculated from Eq.~ (\ref{eta-3}) differ approximately by the factor of two.
As we will show below, this moderate effect of the quantum suppression roots to the fact that $\chi$ remains relatively small in that part of the plasma which gives the most significant contribution into radiation and therefore determines the value of the conversion efficiency.
In this interval of parameters, $a_0\simeq a_\mathrm{cr}\simeq a_\mathrm{HB}$, so that the solution to Eq.~(\ref{a0-gam-g}) considerably deviates from its low-field, $a_0\ll a_\mathrm{cr},~a_\mathrm{HB}$, and the high-field $a_0\gg a_\mathrm{cr},~a_\mathrm{HB}$ asymptotics. 
For the supergaussian pulse $F(\rho,\tau)=\exp(-\rho^2-\tau^4)$ as used in \cite{liseykina-njp16,poprz-njp19} the low-field asymptotic of $\eta_\mathrm{rad},$ Eq.~(\ref{eta-3}), is given by
\beq
\eta_\mathrm{rad}\approx 0.20g(\chi)\xi(a_0)a_0^3~.
\label{lowfield}
\eeq
In the opposite, strong-field limit the function $g(\chi)$ slips out, and $\eta_\mathrm{rad}$ tends to a constant $<1$ as it is in the  classical case (cf. Eq.~(29) in \cite{poprz-njp19}).
This limit is however nonphysical within the considered model, as it can only be reached at $\chi>1$ when the electron motion is essentially non-classical, so that a full quantum mechanical treatment is required.
Thus in the following we focus on the intensity interval $a_0\approx 300 - 700$ where the simple low-field asymptotic, Eq.~(\ref{lowfield}), does not apply while the semiclassical description remains relevant.

We compare predictions of this analytic model to the results of 3D PIC simulations (see also \cite{liseykina-njp16,poprz-njp19} for details of the numerical set-up) which describe the interaction of a laser pulse with a plasma of thickness $D=15\lambda_0$, where $\lambda_0=800$~nm corresponding to a Ti:Sapphire laser and initial density $n_0=90n_c=1.55\cdot 10^{23}$~cm$^{-3}$. 
The supergaussian laser pulse with the space-time envelope $F=\exp(-\rho^2-\tau^4)$ is introduced via the time-dependent boundary condition at the surfaces of the numerical box in a way that at the waist plane $x=0$ coincident with the initial position of the left boundary of the plasma target:
\beq
\veca(r,x=0,t)=a_0\bigl[{\bf y}\cos(\omega t)+{\bf z}\sin(\omega t)\bigr]\mathrm{e}^{-(r/r_0)^4-(ct/r_L)^4}~.
\label{a-rxt}
\eeq
Here $r=\sqrt{y^2+z^2}$, $r_0=3.8\lambda$, $r_L=3.0\lambda$ and duration (full-width-half-maximum of the intensity profile) 14.6~fs.
The numerical box had a ~$[40\times 25\times 25]\lambda^3$ size, with 40 grid cells per $\lambda$ in each direction and 125 particles per cell 
for each species. The simulations were performed on $5000\div 10000$ cores of the JURECA Cluster Module at NIC (J\"ulich, Germany).
The code has been modified by introducing the quantum  factor, Eq.~(\ref{g-chi}), in the expression for the radiation friction force. To that end the quantum parameter $\chi,$ Eq.~(\ref{chi}), was calculated at each time step by taking the values of electric and magnetic fields at the position of each macroparticle. 
The value of $\eta_\mathrm{rad}$ was found as the ratio 
\beq
\eta_\mathrm{rad}=\frac{\int dt\int n_e(\vecr,t)~\vecF(\vecr,t)\cdot\vecv(\vecr,t) d^3r}{U_L}
\label{int}
\eeq
with $\vecF$ being the RF force 
corrected by the factor $g(\chi)$ as described above.

The values of $\eta_\mathrm{rad}$ extracted from the simulation are shown on Fig.~\ref{fig1}b) by triangles and diamonds for the classical and quantum cases, correspondingly.
The analytic modeling agrees with the numerical data only qualitatively.
To gain a deeper understanding of the quantum effect the conversion efficiency and to provide a more informative comparison between the theory and the PIC results we introduce the suppression factor defined as
\beq
\mathrm{SF}=\frac{\eta_\mathrm{rad}}{\eta_\mathrm{rad}^{\rm (cl)}}~,
\label{SF}
\eeq
where $\eta_\mathrm{rad}^{\rm (cl)}$ is calculated with $g(\chi)=1$ in (\ref{P-g}).
Results shown in Fig.~\ref{fig1} demonstrate the non monotonic behavior of this factor as calculated from the theory.
The same function extracted from the PIC data behaves qualitatively similar with a loosely defined minimum around $a_0\approx 450$.
Note that the accuracy of the numerical extraction of $\eta_\mathrm{rad}$ from results of the PIC simulation is close to 1\%.
This establishes the lower limit of the conversion efficiency which can be reliably found from the data.
As a result, the values of $\eta_\mathrm{rad}$ for $a_0=350$ and $g=1$ and $a_0=350,~400$ and $g=g(\chi)$ bear a $\sim 100\%$ uncertainty.
However, the mere fact that $g(\chi)\to 1$ in the low-intensity limit in combination with a clear growth of the numerically found $\mathrm{SF}$ at $a_0>500$ makes the existence of the minimum apparent.

\begin{figure*}
\centering
\includegraphics[width=0.45\textwidth]{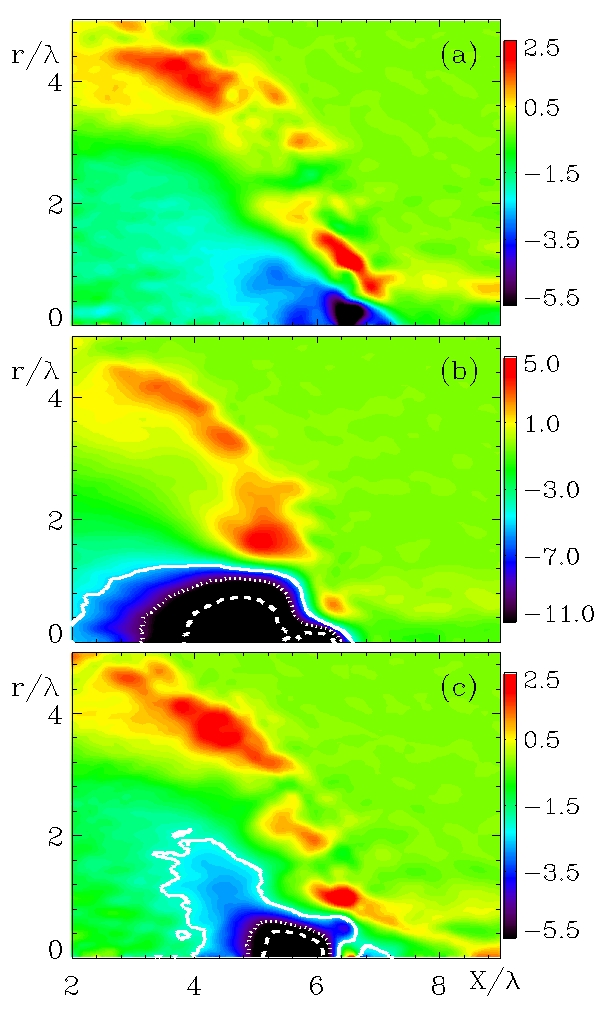}
\includegraphics[width=0.45\textwidth]{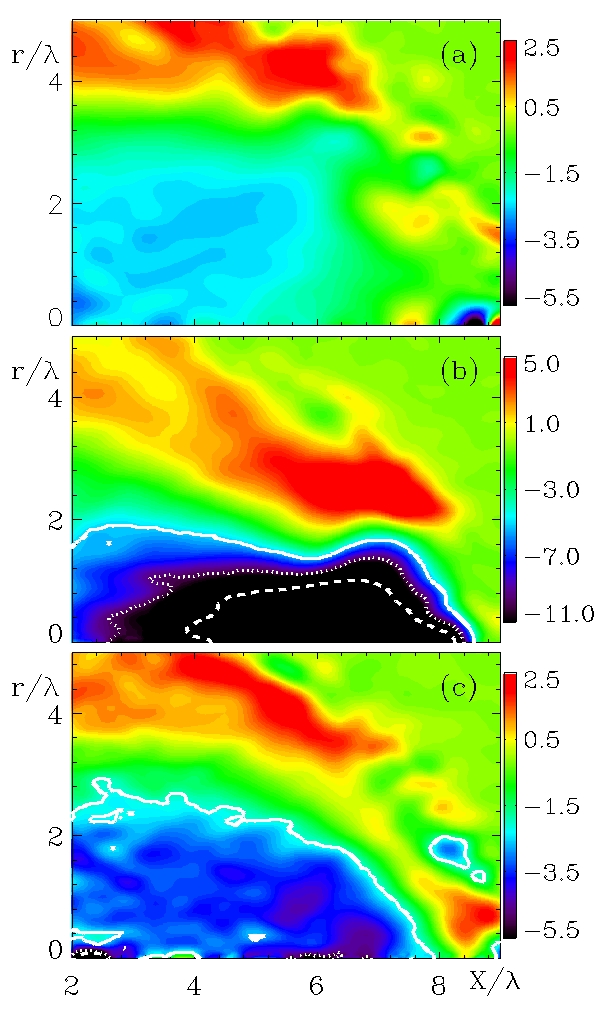}
\caption{Distributions in the longitudinal magnetic field $B_x/B_0$ in the $(x,r)$ plane for $a_0=400$ (left column) and $a_0=500$ (right column) extracted from PIC simulations without the RF force (a), with the RF force for $g=1$ (b) and $g$-factor (\ref{g-chi}) (c).
All distributions are taken at $t=32\cdot 2\pi/\omega$ after the start of the interaction, 
i.e. when the interaction with the laser pulse is over. The white lines on frames (b) and (c) denote 
the levels of the magnetic field strength $[-15,-10,-5]$ and $[-7.5,-5,-2.5],$ respectively.
}
\label{fig2} 
\end{figure*}

To envisage the quantum effect on the generation of extreme magnetic fields via the IFE, we plot on Fig.~\ref{fig2} the distributions of $B_x$ in the plane containing the propagation axis $x$.
The comparison is made for the cases when the quantum factor is taken into account, quantum effects are discarded $(g=1)$, but the RF force is accounted for classically (in the same way as in \cite{liseykina-njp16,poprz-njp19}) and, finally, the RF force is discarded.
It is clearly seen that the quantum effects considerably suppress both the amplitude of the magnetic field and the size of the 
region
where this field exist.
For $a_0=400$ the ratio of the magnetic field amplitudes $B_\mathrm{xm}(g\ne 1)/B_\mathrm{xm}(g=1)\approx 0.5$ and for $a_0=500$ it is close to 0.35.
This results agree qualitatively with Eq.~(\ref{Bmax}) which establishes the linear proportionality of the magnetic field amplitude to the conversion efficiency.
The ratio of magnetic energies calculated in the part of the plasma where the longitudinal magnetic field is present appears $0.07$ for both values of intensity present in Fig.~\ref{fig2}.

\begin{figure*}
\centering
\includegraphics[width=0.9\textwidth]{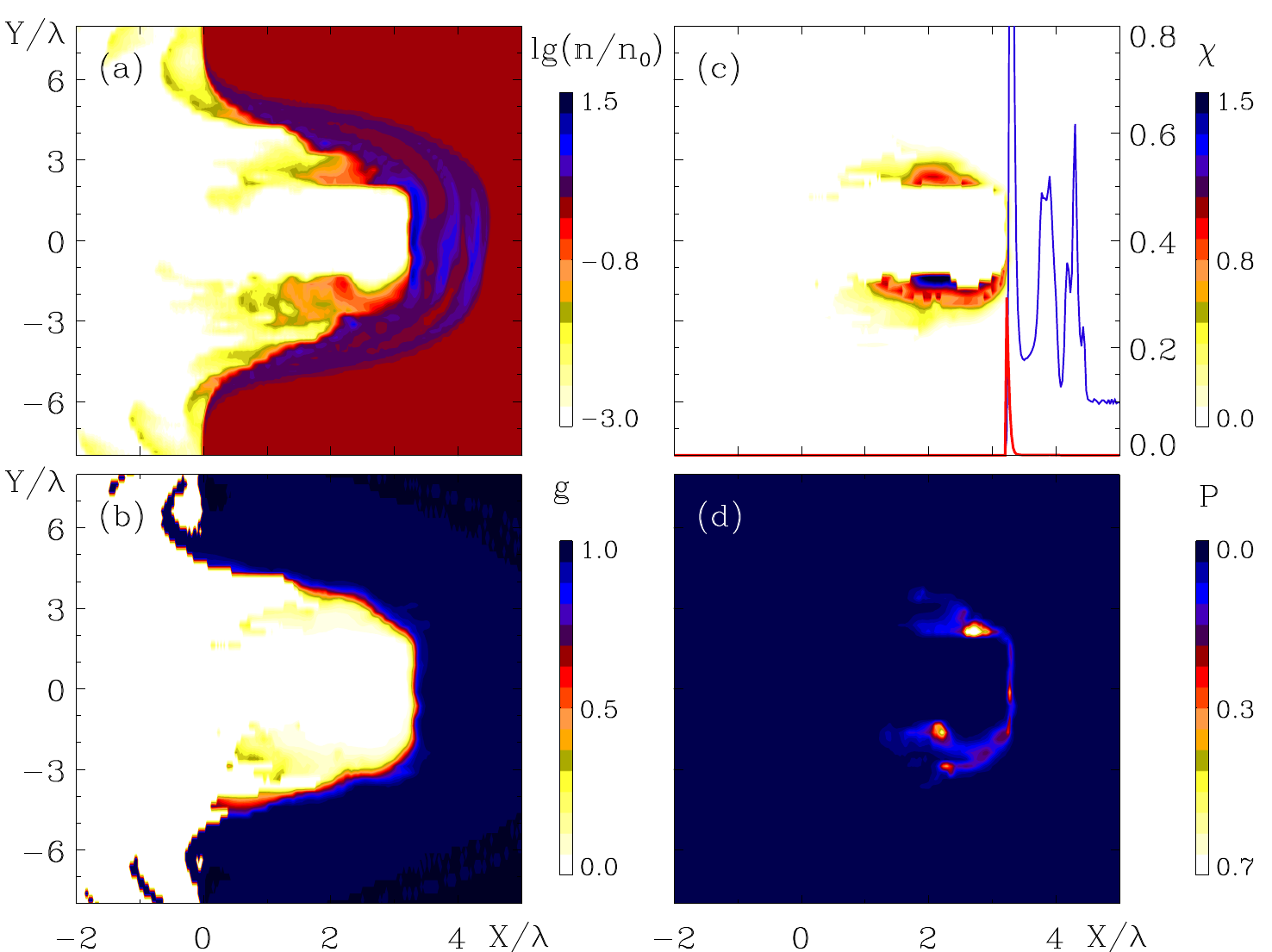}
\caption{Distributions in $(x,y)$ plane for the electron density $n/n_0$ (a), the $g$-factor (b), the $\chi$ parameter (c) and the radiated power (in arbitrary units) (d) for $a_0=500$ and $t=10\cdot 2\pi/\omega.$ 
Panel (c) also shows the cuts of the electron density (blue line relative value, $n/n_0\times 1/10$) and parameter $\chi$ (red line, absolute value) taken at $y=0$, i.e. on the axis of the laser pulse.
Note that although the white regions of the plot of panel (b) formally correspond to $g\to 0$, their contribution into the radiation power and consequently into the value of $\eta_\mathrm{rad}$ is negligible owing to the negligibly small concentration of electrons there.}  
\label{fig3} 
\end{figure*}

\section{Discussion\label{sec:discussion}}

Our numerical and analytic results 
show that the effect of quantum recoil suppressing radiation of high-energy photons leads to a considerable reduction in the conversion efficiency and in the maximal value of the magnetic field excited via the inverse Faraday effect.
The theory predicts the quantum suppression factor to reach the value of $\approx 0.7$ already at $a_0\approx 250$.
This result looks natural taking into account that in the limit of low intensities the hole boring velocity is small so that $\xi\approx\xi_0$ and the RF effect on the electron trajectory 
is negligible so that $\gamma\approx a_0$.
Then $SF\approx g(\chi)$ with $g(\chi=0.1)\approx 0.7$ at $a_0=250$.
The behavior of the suppression factor, Eq. (\ref{SF}), at higher intensities with a minimum at $a_0\approx 500$ looks from the first glance rather counterintuitive.
However, qualitatively it agrees with the PIC results which also demonstrate a minimum localized approximately at $a_0=350-400$ followed by a clear growth at higher intensities.
The value of this minimum $\mathrm{SF}\approx 0.35$ is 
smaller than that predicted by the model.

The very existence of the minimum is not surprising for the applied model which predicts $\eta_\mathrm{rad}=\mathrm{const}$ for $a_0\to\infty$.
Its actual position and value depend on the values of $a_\mathrm{cr}$ and $a_\mathrm{HB}.$
With the increasing intensity the $\gamma$ factor of electrons grows first linearly with $a_0$ and then slower, according to Eq.~(\ref{a0-gam-g}).
This leads to the corresponding growth of the quantum parameter according to Eq.~(\ref{chi-xi}).
At the same time, the hole boring velocity grows almost linearly with $a_0$, according to Eq.~(\ref{vHB}), as long as $a_0<a_\mathrm{HB}$ which leads to the reduction in the values of $\xi$ and $\gamma$ entering Eq.~(\ref{chi-xi}). 
This reduction results in the saturation of the quantum effects in the interval $a_\mathrm{cr}<a_0<a_\mathrm{HB}$.
The growth of $\mathrm{SF}$ at high intensities is the joint result of the longitudinal acceleration of the plasma and the freezing of the relativistic $\gamma$-factor described by the Zeldovich model.
This conclusion should not make a wrong impression that the quantum effects disappear at ultrahigh intensities. 
In fact, at the initial stage of the interaction, before a high value of $v_\mathrm{HB}$ can be achieved, the quantum parameter $\chi$ will reach high values bringing the electron dynamics beyond the classical description. 

The two threshold values $a_\mathrm{cr}$ and $a_\mathrm{HB}$ appeared numerically close in our calculation.
This will not necessarily be the case for all realistic interaction schemes supporting the hole boring regime.
For $a_{\rm HB}<a_{\rm cr}$ the plasma will be easier accelerated up to relativistic velocities, suppressing both the radiation reaction and the quantum effects. 
In this limit, our approach will remain valid, and the conversion efficiency is expected to be smaller than was found here.
In the opposite limit $a_{\rm HB}>a_{\rm cr}$, the hole boring acceleration is slow, and the quantum phenomena will significantly affect the radiation power already at lower intensities pushing the interaction scenario towards the deep quantum regime and making the semiclassical approach used for our calculations inapplicable.

Note that emission of high energy photons proceeds predominantly in a very thin plasma layer near the plasma-vacuum border where the laser pulse is being reflected.
This is illustrated in Fig.~\ref{fig3} where the plasma density, the value of the $g$-factor and the radiation power are shown at a time instant $t=10\cdot 2\pi/\omega$ when the radiation power is close to its maximum.
These distributions help to infer the effective value of the $g$-factor which appears, for $a_0=500$, on the level $g\sim 0.3 - 0.7$.
This corresponds to values of $\chi\approx 0.2$ unexpectedly small for such intensities, which agree however with the 1D distribution of this parameter shown on Fig.\ref{fig3}~(c).
Accurate inspection of the 2D distribution in $\chi$ presented on the same panel shows much higher values of $\chi\simeq 1$.
The semiclassical theory we employ here is invalid for the description of radiation in such essentially quantum regime.
However, these high values of $\chi$ and correspondingly small values of the $g$--factor are achieved predominantly for the electron bunches accelerated by the longitudinal field generated due to the charge separation move with a sufficiently large negative $v_x$.
Such electrons strongly radiate, but their concentration remains low, making their contribution into the full radiation power small.
In principle these electrons could make a considerable and even dominant contributions into the radiation power, owing to the factor $(1-v_x/c)^2\approx 4$ in Eq.~(\ref{P-1}), but the detailed analysis \cite{poprz-njp19} shows that this does not happen for the parameters of our calculation.

In conclusion, we have studied possible effects of the quantum recoil on the generation of ultrahigh magnetic fields through the inverse Faraday effect in a dense plasma irradiated by a short intense circularly polarized infrared laser pulse.
The quantum effect was accounted for by the factor $g(\chi)$ in the radiation power, both in the analytic model and in the PIC simulation, while the electron motion still has been treated classically.
This limits our results by intensities $I_L<10^{25}\Wcm$ where $\chi<1$ in that part of the target which efficiently emits radiation.
We show a considerable suppression both in the conversion efficiency $\eta_\mathrm{rad}$ and in the magnetic field amplitude $B_\mathrm{xm}$ and, most importantly, we demonstrate that this suppression is non monotonic with intensity, reaching a maximum at  $I_L\simeq 10^{24}\Wcm.$

\section*{Acknowledgment}
Authors acknowledge useful discussions on quantum radiation reaction with S.~S.~Bulanov, A.~Di~Piazza, A.~M.~Fedotov and M.~Tamburini. 
The simulations were performed using the computing resources granted by the John von Neumann-Institut f\"ur Computing (Research Center J\"ulich) under the project HRO04. 
 T.V.L. acknowledges the support by the state contract with ICMMG SB RAS (0315-2019-0009) and by the Ministry of Science
and Higher Education of the Russian Federation (agreement 075-03-2020-223/3
within FSSF-2020-0018) in the part related to the analysis of the numerical
results. S.V.P. acknowledges support from the Russian Science Foundation through grant No 20-12-00077 in the part of the work devoted to the design of the analytic model.

\bibliographystyle{iopart-num.bst}

\hyphenation{Post-Script Sprin-ger}
\providecommand{\newblock}{}
\begin{thebibliography}{10}
\expandafter\ifx\csname url\endcsname\relax
  \def\url#1{{\tt #1}}\fi
\expandafter\ifx\csname urlprefix\endcsname\relax\def\urlprefix{URL }\fi
\providecommand{\eprint}[2][]{\url{#2}}

\bibitem{danson-hpl15}
Danson C, Hillier D, Hopps N and Neely D 2015 {\em High Power Laser Science and
  Engineering\/} {\bf 3} e3 ISSN 2052-3289

\bibitem{light-f}
Cartlidge E 2018 {\em Science\/} {\bf 359} 382--385
  \urlprefix\url{https://science.sciencemag.org/content/359/6374/382}

\bibitem{ELI}
Chambaret J, Chekhlov O, Ch{\'e}riaux G, Collier J, Dabu R, Dombi P, Dunne A,
  Ertel K, Georges P, Hebling J, Hein J, Hernandez-Gomez C, Hooker C, Karsch S,
  Korn G, Krausz F, {Le Blanc} C, Major Z, Mathieu F, Metzger T, Mourou G,
  Nickles P, Osvay K, Rus B, Sandner W, Szab{\'o} G, Ursescu D and Varj{\'u} K
  2010 Extreme light infrastructure: Architecture and major challenges {\em
  Solid State Lasers and Amplifiers IV, and High-Power Lasers\/} Proceedings of
  SPIE - The International Society for Optical Engineering ISBN 9780819481948
  solid State Lasers and Amplifiers IV, and High-Power Lasers ; Conference
  date: 12-04-2010 Through 16-04-2010

\bibitem{P3-ELI}
Weber S, Bechet S, Borneis S, Brabec L, Bučka M, Chacon-Golcher E, Ciappina M,
  DeMarco M, Fajstavr A, Falk K, Garcia E~R, Grosz J, Gu Y~J, Hernandez J~C,
  Holec M, Janečka P, Jantač M, Jirka M, Kadlecova H, Khikhlukha D, Klimo O,
  Korn G, Kramer D, Kumar D, Lastovička T, Lutoslawski P, Morejon L,
  Olšovcová V, Rajdl M, Renner O, Rus B, Singh S, Šmid M, Sokol M, Versaci
  R, Vrána R, Vranic M, Vyskočil J, Wolf A and Yu Q 2017 {\em Matter and
  Radiation at Extremes\/} {\bf 2} 149 -- 176 ISSN 2468-080X
  \urlprefix\url{http://www.sciencedirect.com/science/article/pii/S2468080X17300171}

\bibitem{apollon}
Papadopoulos D, Zou J, Le~Blanc C, Chériaux G, Georges P, Druon F, Mennerat G,
  Ramirez P, Martin L, Fréneaux A and et~al 2016 {\em High Power Laser Science
  and Engineering\/} {\bf 4} e34

\bibitem{J-Karen}
Pirozhkov A~S, Fukuda Y, Nishiuchi M, Kiriyama H, Sagisaka A, Ogura K, Mori M,
  Kishimoto M, Sakaki H, Dover N~P, Kondo K, Nakanii N, Huang K, Kanasaki M,
  Kondo K and Kando M 2017 {\em Opt. Express\/} {\bf 25} 20486--20501
  \urlprefix\url{http://www.opticsexpress.org/abstract.cfm?URI=oe-25-17-20486}

\bibitem{SULF}
Guo Z, Yu L, Wang J, Wang C, Liu Y, Gan Z, Li W, Leng Y, Liang X and Li R 2018
  {\em Opt. Express\/} {\bf 26} 26776--26786
  \urlprefix\url{http://www.opticsexpress.org/abstract.cfm?URI=oe-26-20-26776}

\bibitem{SULF-1}
Li W, Gan Z, Yu L, Wang C, Liu Y, Guo Z, Xu L, Xu M, Hang Y, Xu Y, Wang J,
  Huang P, Cao H, Yao B, Zhang X, Chen L, Tang Y, Li S, Liu X, Li S, He M, Yin
  D, Liang X, Leng Y, Li R and Xu Z 2018 {\em Opt. Lett.\/} {\bf 43} 5681--5684
  \urlprefix\url{http://ol.osa.org/abstract.cfm?URI=ol-43-22-5681}

\bibitem{SULF-2}
Gan Z, Yu L, Li S, Wang C, Liang X, Liu Y, Li W, Guo Z, Fan Z, Yuan X, Xu L,
  Liu Z, Xu Y, Lu J, Lu H, Yin D, Leng Y, Li R and Xu Z 2017 {\em Opt.
  Express\/} {\bf 25} 5169--5178
  \urlprefix\url{http://www.opticsexpress.org/abstract.cfm?URI=oe-25-5-5169}

\bibitem{APRI}
Sung J~H, Lee H~W, Yoo J~Y, Yoon J~W, Lee C~W, Yang J~M, Son Y~J, Jang Y~H, Lee
  S~K and Nam C~H 2017 {\em Opt. Lett.\/} {\bf 42} 2058--2061
  \urlprefix\url{http://ol.osa.org/abstract.cfm?URI=ol-42-11-2058}

\bibitem{CAEP}
Zeng X, Zhou K, Zuo Y, Zhu Q, Su J, Wang X, Wang X, Huang X, Jiang X, Jiang D,
  Guo Y, Xie N, Zhou S, Wu Z, Mu J, Peng H and Jing F 2017 {\em Opt. Lett.\/}
  {\bf 42} 2014--2017
  \urlprefix\url{http://ol.osa.org/abstract.cfm?URI=ol-42-10-2014}

\bibitem{XCELS}
Bashinov A, Gonoskov A, Kim A, Mourou G and Sergeev A 2014 {\em The European
  Physical Journal Special Topics\/} {\bf 223} 1105--1112
  \urlprefix\url{https://doi.org/10.1140/epjst/e2014-02161-7}

\bibitem{bulanov-rmp06}
Mourou G~A, Tajima T and Bulanov S~V 2006 {\em Rev. Mod. Phys.\/} {\bf 78}(2)
  309--371 \urlprefix\url{https://link.aps.org/doi/10.1103/RevModPhys.78.309}

\bibitem{dipiazza-rmp12}
Di~Piazza A, M\"uller C, Hatsagortsyan K~Z and Keitel C~H 2012 {\em Rev. Mod.
  Phys.\/} {\bf 84}(3) 1177--1228
  \urlprefix\url{http://link.aps.org/doi/10.1103/RevModPhys.84.1177}

\bibitem{fedotov-cp15}
Narozhny N~B and Fedotov A~M 2015 {\em Contemporary Physics\/} {\bf 56}
  249--268 (\textit{Preprint}
  \eprint{https://doi.org/10.1080/00107514.2015.1028768})
  \urlprefix\url{https://doi.org/10.1080/00107514.2015.1028768}

\bibitem{landau-lifshitz-RR}
Landau L~D and Lifshitz E~M 1975 {\em The Classical Theory of Fields\/}
  (Elsevier, Oxford) chap~76 2nd ed

\bibitem{cole-prx18}
Cole J~M, Behm K~T, Gerstmayr E, Blackburn T~G, Wood J~C, Baird C~D, Duff M~J,
  Harvey C, Ilderton A, Joglekar A~S, Krushelnick K, Kuschel S, Marklund M,
  McKenna P, Murphy C~D, Poder K, Ridgers C~P, Samarin G~M, Sarri G, Symes D~R,
  Thomas A~G~R, Warwick J, Zepf M, Najmudin Z and Mangles S~P~D 2018 {\em Phys.
  Rev. X\/} {\bf 8}(1) 011020
  \urlprefix\url{https://link.aps.org/doi/10.1103/PhysRevX.8.011020}

\bibitem{poder-prx18}
Poder K, Tamburini M, Sarri G, Di~Piazza A, Kuschel S, Baird C~D, Behm K,
  Bohlen S, Cole J~M, Corvan D~J, Duff M, Gerstmayr E, Keitel C~H, Krushelnick
  K, Mangles S~P~D, McKenna P, Murphy C~D, Najmudin Z, Ridgers C~P, Samarin
  G~M, Symes D~R, Thomas A~G~R, Warwick J and Zepf M 2018 {\em Phys. Rev. X\/}
  {\bf 8}(3) 031004
  \urlprefix\url{https://link.aps.org/doi/10.1103/PhysRevX.8.031004}

\bibitem{macchi-p18}
Macchi A 2018 {\em Physics\/} {\bf 11} 13
  \urlprefix\url{https://physics.aps.org/articles/v11/13}

\bibitem{kirk-ppcf09}
Kirk J, Bell A~R and Arka I 2009 {\em Plasma Physics and Controlled Fusion\/}
  {\bf 51} 085008 \urlprefix\url{https://doi.org/10.1088/0741-3335/51/8/085008}

\bibitem{wistisen-natcomm18}
Wistisen T~N, Di~Piazza A, Knudsen H~V and Uggerh{\o}j U~I 2018 {\em Nature
  Communications\/} {\bf 9}(1) 795
  \urlprefix\url{https://doi.org/10.1038/s41467-018-03165-4}

\bibitem{blackburn-pop18}
Blackburn T~G, Seipt D, Bulanov S~S and Marklund M 2018 {\em Physics of
  Plasmas\/} {\bf 25} 083108 \urlprefix\url{https://doi.org/10.1063/1.5037967}

\bibitem{dipiazza-pra18}
Di~Piazza A, Tamburini M, Meuren S and Keitel C~H 2018 {\em Phys. Rev. A\/}
  {\bf 98}(1) 012134
  \urlprefix\url{https://link.aps.org/doi/10.1103/PhysRevA.98.012134}

\bibitem{ildertonPRA19}
Ilderton A, King B and Seipt D 2019 {\em Phys. Rev. A\/} {\bf 99}(4) 042121
  \urlprefix\url{https://link.aps.org/doi/10.1103/PhysRevA.99.042121}

\bibitem{aleksandrovPRD19}
Aleksandrov I~A, Plunien G and Shabaev V~M 2019 {\em Phys. Rev. D\/} {\bf
  100}(11) 116003
  \urlprefix\url{https://link.aps.org/doi/10.1103/PhysRevD.100.116003}

\bibitem{dipiazza-pra19}
Di~Piazza A, Tamburini M, Meuren S and Keitel C~H 2019 {\em Phys. Rev. A\/}
  {\bf 99}(2) 022125
  \urlprefix\url{https://link.aps.org/doi/10.1103/PhysRevA.99.022125}

\bibitem{seiptPPCF19}
Seipt D and Thomas A~G~R 2019 {\em Plasma Physics and Controlled Fusion\/} {\bf
  61} 074005
  \urlprefix\url{https://iopscience.iop.org/article/10.1088/1361-6587/ab1e77}

\bibitem{wistisenPRR19}
Wistisen T~N, Di~Piazza A, Nielsen C~F, S\o{}rensen A~H and Uggerh\o{}j U~I
  (CERN NA63) 2019 {\em Phys. Rev. Research\/} {\bf 1}(3) 033014
  \urlprefix\url{https://link.aps.org/doi/10.1103/PhysRevResearch.1.033014}

\bibitem{dinuPRD20}
Dinu V and Torgrimsson G 2020 {\em Phys. Rev. D\/} {\bf 102}(1) 016018
  \urlprefix\url{https://link.aps.org/doi/10.1103/PhysRevD.102.016018}

\bibitem{liseykina-njp16}
Liseykina T~V, Popruzhenko S~V and Macchi A 2016 {\em New Journal of Physics\/}
  {\bf 18} 072001

\bibitem{poprz-njp19}
Popruzhenko S~V, Liseykina T~V and A M 2019 {\em New Journal of Physics\/} {\bf
  21} 033009 \urlprefix\url{http://iopscience.iop.org/10.1088/1367-2630/ab0119}

\bibitem{zeld-spu75}
Zel'dovich Y~B 1975 {\em Soviet Physics Uspekhi\/} {\bf 18} 79
  \urlprefix\url{http://stacks.iop.org/0038-5670/18/i=2/a=R01}

\bibitem{macchi-prl05}
Macchi A, Cattani F, Liseykina T~V and Cornolti F 2005 {\em Phys. Rev. Lett.\/}
  {\bf 94}(16) 165003
  \urlprefix\url{http://link.aps.org/doi/10.1103/PhysRevLett.94.165003}

\bibitem{robinson-ppcf09}
Robinson A~P~L, Gibbon P, Zepf M, Kar S, Evans R~G and Bellei C 2009 {\em
  Plasma Physics and Controlled Fusion\/} {\bf 51} 024004
  \urlprefix\url{https://doi.org/10.1088/0741-3335/F51/2/024004}

\bibitem{ritus-jslr85}
Ritus V~I 1985 {\em Journal of Soviet Laser Research\/} {\bf 6} 497--617 ISSN
  1573-8760 \urlprefix\url{https://doi.org/10.1007/BF01120220}

\bibitem{ritus}
Ritus V~I 1979 {\em Moscow Izdatel Nauka AN SSR Fizicheskii Institut Trudy\/}
  {\bf 111} 5--151

\bibitem{niel-ppcf18}
Niel F, Riconda C, Amiranoff F, Lobet M, Derouillat J, P{\'{e}}rez F, Vinci T
  and Grech M 2018 {\em Plasma Physics and Controlled Fusion\/} {\bf 60} 094002
  \urlprefix\url{https://doi.org/10.1088/1361-6587/aace22}

\bibitem{pitaevskiiJETP61}
Pitaevskii L~P 1961 {\em Sov. Phys. JETP\/} {\bf 12} 1008--1013

\bibitem{haines-prl01}
Haines M~G 2001 {\em Phys. Rev. Lett.\/} {\bf 87} 135005

\bibitem{sauter}
Sauter F 1931 {\em Zeitschrift f\"ur Physik\/} {\bf 69} 742--764

\bibitem{heisenberg}
Heisenberg W and Euler H 1935 {\em Zeitschrift f\"ur Physik\/} {\bf 98}
  714--732

\bibitem{schwinger}
Schwinger J 1951 {\em Phys. Rev.\/} {\bf 82}(5) 664--679
  \urlprefix\url{https://link.aps.org/doi/10.1103/PhysRev.82.664}

\bibitem{thomas-prx12}
Thomas A~G~R, Ridgers C~P, Bulanov S~S, Griffin B~J and Mangles S~P~D 2012 {\em
  Phys. Rev. X\/} {\bf 2}(4) 041004
  \urlprefix\url{https://link.aps.org/doi/10.1103/PhysRevX.2.041004}

\bibitem{zhang-njp15}
Zhang P, Ridgers C~P and Thomas A~G~R 2015 {\em New Journal of Physics\/} {\bf
  17} 043051 \urlprefix\url{https://doi.org/10.1088/1367-2630/17/4/043051}

\end{thebibliography}

\end{document}